\documentclass{iau}
\usepackage{graphicx}
\usepackage[usenames, dvipsnames]{color}
\usepackage{natbib}

\usepackage{hyperref}
\hypersetup{
    final=true,
    pageanchor=true,
    colorlinks=true,
    breaklinks=true,
    linkcolor=blue,
    citecolor=blue,
    urlcolor=blue,
    pdfpagemode=UseNone,
    pdftitle={High-resolution spectroscopy of a giant solar filament},
    pdfauthor={C. Kuckein, C. Denker and M. Verma},
    pdfsubject={Solar Physics},
    pdfkeywords={Sun: filaments, Sun: chromosphere, Sun: photosphere, techniques:
spectroscopic, techniques: high angular resolution}}

\newcommand*\degr{\ensuremath{^\circ}}

\newcommand*\arcsec{\ensuremath{^{\prime\prime}}}

\title[High-resolution spectroscopy of a giant solar filament] 
{High-resolution spectroscopy of a giant solar filament}

\author[Christoph Kuckein, Carsten Denker and Meetu Verma]   
{Christoph Kuckein, Carsten Denker \and Meetu Verma}

\affiliation{Leibniz-Institut f\"ur Astrophysik (AIP),  
An der Sternwarte 16, 14482, Potsdam, Germany  \\
email: {\tt ckuckein@aip.de}}

\pubyear{2013}
\volume{300}
\pagerange{1--2}
\jname{Nature of Prominences and their Role in Space Weather}
\editors{B.\ Schmieder, J.-M.\ Malherbe \& S.\ Wu, eds.}

\begin{document}

\maketitle

\begin{abstract}
High-resolution spectra of a giant solar quiescent filament were taken with the Echelle spectrograph at the Vacuum Tower
Telescope (VTT; Tenerife, Spain). A mosaic of various spectroheliograms (H$\alpha$, H$\alpha \pm 0.5$~\AA\ and Na
D$_2$) were chosen to examine the filament at different heights in the solar atmosphere. In addition, full-disk images
(He \textsc{i} 10830~\AA\ and Ca \textsc{ii} K) of the Chromspheric Telescope and full-disk
magnetograms of the Helioseismic and Magnetic Imager were used to complement the
spectra. Preliminary results are shown of this filament, which had extremely large linear
dimensions ($\sim$740\arcsec) and was observed in November 2011 while it traversed the northern solar hemisphere.

\end{abstract}

\firstsection 
\section{Introduction}
Filaments are large structures observed in the solar corona or chromosphere. On the disk, they are seen as
elongated dark features whereas above the limb they appear bright against the dark background and are called
prominences. They have been observed for many centuries and are largely classified into two groups: quiescent (QS)
filaments and active region (AR) filaments. The so-called polar crown filaments are long QS filaments that lie at
high latitudes ($> 45^\circ$) and usually form a crown around the pole \citep[][]{cartledge96}.
The observed filament in this work does not fit into the scheme of a polar crown filament. It is rather a QS
filament which is particularly interesting owing to its large linear dimensions and its location across the
northern solar disk. In the literature, only a few examples of giant QS filaments are found 
\citep[e.g.,][]{yazev88}.

\section{Observations}
The ground-based observations of the filament were acquired in 2011 November 15 with the
Echelle spectrograph at the VTT. The good seeing conditions
made it possible to scan with the slit the whole filament along the northern hemisphere of the Sun. The
observing strategy was to divide the filament into ten pieces. For each piece the scanned area was 
$100\arcsec \times 182\arcsec$ and consecutive scans slightly overlapped to assure the continuity of the 
filament and to facilitate the subsequent reconstruction of the mosaic. The whole filament was scanned
between 11:38\,UT and 13:11\,UT, starting at heliographic coordinates (59\degr~E, 48\degr~N) and ending at
(14\degr~W, 18\degr~N). The filament extends roughly 740\arcsec\ (536~Mm) from solar East to West. 

Two CCD cameras were mounted at the Echelle spectrograph to acquire two different spectral regions. The first one 
was centered at the chromospheric H$\alpha$ line at 6562.8~\AA\ and spanned a spectral range of 8~\AA. 
The second region was centered at the Na D$_2$ line at 5889.9~\AA\ and covered a spectral range of 7~\AA.

In addition, full-disk images of the Chromospheric Telescope \citep[ChroTel;][]{bethge11} 
attached to the VTT building were used for this study.
ChroTel acquires images at several wavelengths (H$\alpha$, He\,\textsc{i} 10830~\AA\ and Ca\,\textsc{ii} K)
with a cadence of 3~min.  Furthermore, full-disk magnetograms of the Helioseismic and Magnetic Imager instrument
 \citep[HMI;][]{HMI} were used to complement the spectra. The magnetograms confirm that the
filament
lies on top of the polarity inversion line (PIL). Positive (negative) polarity is above (below) the
filament as shown in Fig. \ref{fig1}.  

\section{Results}
\begin{figure}[t]
\begin{center}
 \includegraphics[width=\textwidth]{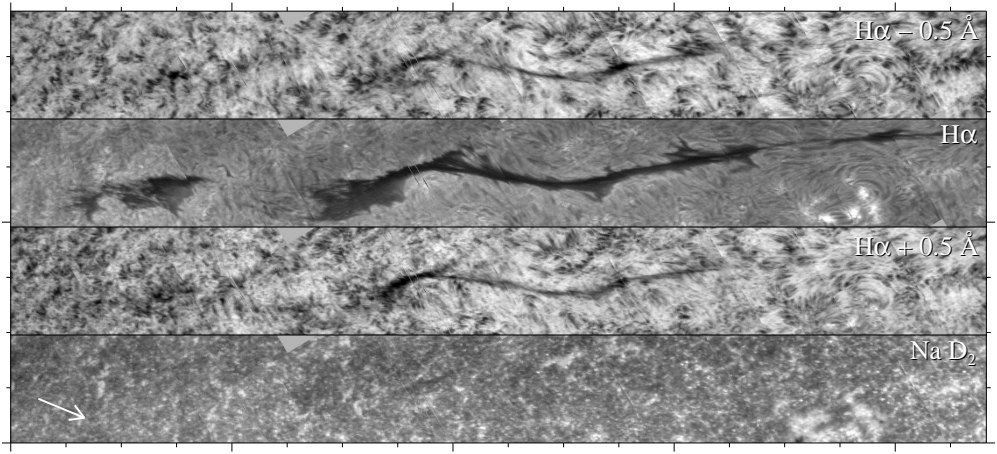} 
 \vspace*{-0.6 cm}
 \caption{The spectroheliograms were assembled from ten partially overlapping slit-reconstructed images:
H$\alpha - 0.5$~\AA, H$\alpha$ line core, H$\alpha + 0.5$~\AA\ and Na D$_2$ (\textit{top to bottom}). On the
right hand side an emerging flux region appears close to the filament. Major tick marks
are separated by 200\arcsec. The white arrow points towards the disk center.  }
   \label{fig1}
\end{center}
\end{figure}

The filament with extraordinary linear dimensions shows a gap on the left hand side in the H$\alpha$ panel of
Fig. \ref{fig1}. A few hours before, plasma was
removed from that side by a coronal mass ejection (CME). However, thin threads appear to link
both ends of the separated filament. Barbs, i.e., groups of thin threads that protrude from the side
of the main body of the filament \citep[e.g.,][]{martin98}, are detected. The spine of the filament is
also seen at H$\alpha \pm 0.5$~\AA. In these areas the H$\alpha$ line is extremely broad. Very little absorption at
the filament can be seen in the Na D$_2$ line core panel.

\begin{acknowledgments}
\noindent CK greatly acknowledges the travel support received from the IAU. CD was supported by grant DE 787/3-1 of the
German Science Foundation (DFG). MV thanks the German Academic Exchange Service (DAAD) for its support in the form of a
PhD scholarship. 
\end{acknowledgments}

\end{document}